\documentclass[lnbip,sechang,a4paper]{svmultln}
\usepackage{amssymb}
\usepackage{graphicx}
\usepackage{caption,subfig}
\usepackage[cmex10]{amsmath}
\usepackage{tikz} 

\usepackage{url}
\urldef{\mailsa}\path|{mdtamzeedislam, bashimaislam}@gmail.com, eunus@cse.buet.ac.bd|
\usepackage[pdfpagelabels,hypertexnames=false,breaklinks=true,bookmarksopen=true,bookmarksopenlevel=2]{hyperref}

\begin{document}

\mainmatter  

\title{A System for Identifying and Visualizing Influential Communities}

%
%
\author{Md Tamzeed Islam, Bashima Islam
\and Mohammed Eunus Ali}  %
\authorrunning{A System for Identifying and Visualizing Influential Communities}

\institute{Department of Computer Science \& Engineering\\
Bangladesh University of Engineering \& Technology\\
Dhaka 1000, Bangladesh
\mailsa}

%
%

\toctitle{A System for Identifying and Visualizing Influential Communities}
\tocauthor{Authors' Instructions}
\maketitle

\begin{abstract}
In this paper, we introduce the concept of influential communities in a co-author network. We term a community as the most influential if the community has the highest influence among all other communities in the entire network. Influence of a community depends on the impact of the contents (e.g., citations of papers) generated by the members of that community. We propose an algorithm to identify the top $K$ influential communities of an online social network. As a working prototype, we develop a visualization system that allows a user to find the top $K$ influential communities from a co-author network. A user can search top $K$ influential communities of particular research fields and our system provides him/her with a visualization of these communities. A user can explore the details of a community, such as authors, citations, and collaborations with other communities.
\end{abstract}

\section{Introduction} \label{into}
Community detection in online social networks has gained enormous importance due to its ubiquitous applications~\cite{newman, himel}. Applications include online marketing, recommendation systems, load balancing, etc. We envision a new set of applications of community detection, that requires identifying the influential communities. Consider the following scenario. An aspiring researcher is looking for the current research trends in the field of data mining. Consequently, she wants to find out about the works of the most\textit{ influential} research group in data mining. In this scenario, she needs to identify the \textit{community} that has the highest influence in data mining area in a co-author network.  
\par
\begin{figure}[!h]
\centering
  \includegraphics[height=1.5in, width=2.5in]{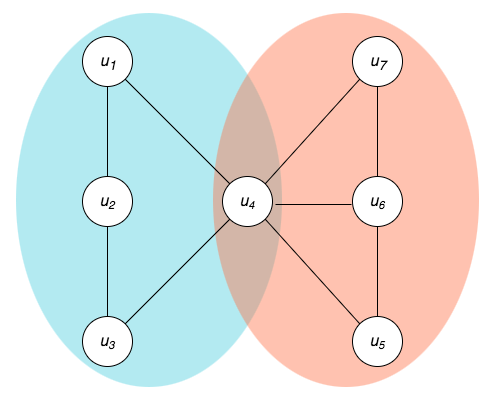}
  \caption{A Co-author Network}
  \label{fig_graph}
  \vspace{-8mm}
\end{figure}

There have been extensive studies on community detection in recent years. Commonly used methods of community detection involve modularity \cite{clauset} \cite{meo} and edge betweenness \cite{new} \cite{conga}. There have been other approaches that formulate the community detection problem in terms of dense region identification, e.g., dense clique detection \cite{abello} \cite{zeng}, minimum cut \cite{newman} and label propagation \cite{rag} based techniques. Recently, Li et al.\cite{rong} proposed a technique to identify \textit{influential} communities, where they only consider the underlying network structure to measure the \textit{influence} of communities. 
\par

In this paper, we define the \textit{influence} of a community by considering the structure, interaction, and the content generated by the community. Our definition of \textit{influential} community is based on the observation that, the influence of a community depends on the \textit{central tendency} (e.g., weighted mean)  of the  impact of contents generated by the members of that community.

\begin{table}[h]
\centering
\caption{Citation of papers by authors of   $\left\lbrace c_{1}, c_{2}\right\rbrace$ }
\label{table_example}
\begin{tabular}{|l|l|l|l|}
\hline
Paper    & Authors & Community & Number of Citations \\ \hline
$p_1$    &   $u_1,u_2$      & $c_1$     & $0$                 \\ \hline
$p_2$    &   $u_1,u_4$     & $c_1$     & $0$                 \\ \hline
$p_3$    &   $u_2,u_3$     & $c_1$     & $3$                 \\ \hline
$p_4$    &   $u_3,u_4$       & $c_1$     & $3$                 \\ \hline
$p_5$    &   $u_1,u_2$       & $c_1$     & $50$                \\ \hline
$p_6$    &     $u_4,u_5,u_6$    & $c_2$     & $9$                 \\ \hline
$p_7$    &     $u_6,u_7$    & $c_2$     & $10$                 \\ \hline
$p_8$    &    $u_4,u_7$     & $c_2$     & $11$                 \\ \hline

\end{tabular}
\end{table}

\par
To illustrate the concept of \textit{influential} community, let us consider a co-author network depicted in Figure~\ref{fig_graph} and Table~\ref{table_example}. There are seven authors   $\left\{{u_1, u_2, ... , u_7}\right\}$ and two communities $\left\{c_1, c_2\right\}$ in this example. The set of authors $\left\{u_1, u_2, u_3, u_4\right\}$ and $\left\{u_4, u_5, u_6, u_7\right\}$  belong to community $c_1$ and $c_2$, respectively. The sets of papers written by the members of $c_1$ and $c_2$ are $\left\{p_1, p_2, p_3, p_4, p_5\right\}$ and $\left\{p_6, p_7, p_8\right\}$, respectively. 

Li et al.~\cite{rong}  define the influential community as the community where the minimum number of citations of an author of that community is the maximum among that of the any other communities in the network. Thus, to qualify for being the influential community, all the papers written by the members of that community should have high number of citations. Hence, the minimum number of citations of papers of a community capture this intuitive idea. In Table~\ref{table_example}, the influence of community $c_1$  and $c_2$ are $0$ and $9$ respectively. However, using minimum number of citations as an influence metric does not reflect the central tendency of the number of citations of papers. Simple statistical measure such as arithmetic mean is not an appropriate definition of central tendency as this is prone to outlier data. From Table~\ref{table_example} the arithmetic mean of the papers written by $c_1$ and $c_2$ are $(0+0+3+3+50)/5 =11.2$ and $(9+10+11)/3= 10$, respectively. Here, the mean of $c_1$ is higher because of high citations of $p_5$. However, the overall tendency of citations of $p_1, p_2, p_3, p_4$ are much lower than that of the papers of $c_2$. The number of citations of $p_5$ does not reflect the true central tendency of the number of citations for $c_1$. One of the popular author-level metrics that measures the bibliometric impact of individual author is h-index \cite{hindex}. We cannot extend this metric for measuring the influence of a community because it is susceptible to the number of publications. For example from Table~\ref{table_example}, we can deduce that the h-index of both $c_1$ and $c_2$ are $3$. This is due to the fact that  $c_2$ has a lower number of publications.

To avoid the above limitations, we introduce a new metric to measure the impact of a paper, where we use the citation count of a paper with a weight factor to capture the central tendency of the number of citations of papers. We define the influence of a community using this new metric. We also propose an algorithm that identifies the top $K$ \textit{influential} communities using  the proposed definitions. 

We develop an interactive visualization system, \textbf{VizCom}, which demonstrates the top $K$  \textit{influential} communities in a co-author network. \textbf{VizCom} assists a user in finding the most \textit{influential} communities of co-authors in specific research fields. A user can get elaborate information about these communities e.g., citations, collaborative communities, collaborating authors between two communities. Along with this information, \textbf{VizCom} provides a user with the details about the authors in these communities, such as affiliations, citations, papers, and co-authors.

In summary, our contributions are as follows:
\begin{itemize}
\item We formally define the \textit{influence} of a community in a co-author network. 
\item We develop an algorithm for identifying top $K$ \textit{influential} communities. 
\item We demonstrate a visualizer for presenting influential communities by using a prototype web application, \textbf{VizCom}. 
\end{itemize}

\section{Finding Top $K$ Influential Communities}
We model the co-author network as a  graph where nodes and edges represent authors and relationships among authors, respectively. For example, in Figure~\ref{fig_graph}  $\left\lbrace u_{1}, u_{2}, ... , u_{7}\right\rbrace$  represent authors and the edges represent the collaboration between the connecting authors. From this co-author network, we first identify communities, and then apply our metrics for identifying influential communities.

\subsection{Community Detection}

Community detection for a co-author network involves grouping of authors who collaborate with each other frequently. We identify a community based on Cluster-Overlap Newman-Girvan Algorithm (CONGA)\cite{conga}. CONGA extends Girvan and Newman's (GN) \cite{new} well-known algorithm for finding a community. GN is based on the betweenness centrality measure. The betweenness $c_{B}(e)$ of an edge $e$ is defined as the number of shortest paths between all pairs of vertices that pass through $e$. In~\cite{new} they consider communities as disjoint structures. However, overlapping communities exist in many real world networks, such as co-author networks. CONGA introduces the concept of vertex betweenness and split betweenness to detect nodes which belong to multiple communities. Vertex betweenness of a vertex $v$ is the total number of shortest paths passing through $v$, and split betweenness of a vertex $v$ is the number of shortest paths passing between the members of two disjoint sets containing all neighbours of $v$. In summary, we can summarize the steps CONGA as follows:

\begin{itemize}
\item Calculate the edge betweenness, vertex betweenness, and split betweenness.
\item Remove the edge with the maximum edge betweenness or split vertex with the maximum split betweenness, given that split betweenness is greater than the maximum edge betweenness.
\item Recalculate the above-mentioned parameters and repeat until no edges remain in the graph.
\end{itemize}

We remove edges and split nodes to get a dendrogram as an end result. By horizontally cutting the dendrogram at an appropriate height, we get the required number of communities. This appropriate height is calculated by using modularity defined by Clauset et al. \cite{clauset}.

\subsection{Influence of Community}

Let $c_{i}$ be a community of $q$ authors  $ \left\lbrace u_{1}, u_{2}, ... , u_{q} \right\rbrace  $. Let us assume that $p_{j}$ be a paper written by some members of $c_{i}$. The set of authors and number of citations of paper $p_{j}$ are defined as $author\left( p_{j}\right) $ and $cite\left( p_{j}\right)$, respectively. We define a function $\Delta c_{i} p_{j}$ that returns 1 if $p_{j}$ is written \emph{only} by the members of community $c_{i}$, which can be defined as follows. 

\begin{equation}
  \Delta c_{i} p_{j}=\left\{
  \begin{array}{@{}ll@{}}
    1, & \text{if } author \left( p_{j} \right) \ \subset \ c_{i} \\
    0, & \text{otherwise}
  \end{array}\right.
\end{equation} 

For example, in Table~\ref{table_example}, $\Delta c_{1} p_{1}=1$ and $\Delta c_{1} p_{8}=0$ as authors $\left\lbrace u_{1}, u_{2}\right\rbrace \subset c_1 $ and $\left\lbrace u_{4}, u_{7}\right\rbrace \not\subset c_1 $.
 
We define the impact of a paper such that, it demonstrates the citation count of the paper and reflects the central tendency of citations of papers written by a particular community. For a paper $p_j$ of community $c_i$,  we multiply $cite\left(p_j \right)$ with a weighted factor $w\left(p_j \right)$. Here, $w\left(p_j \right)$ denotes the probability that the number of citations of a paper written by community $c_i$ is higher than or equal to $cite\left(p_j \right)$. This factor captures the overall central tendency of the citation of the community's papers by reducing the weight of the outlier data.

For two papers $p_l$ and $p_k$ we first define a function $\delta p_l p_k$ as follows.
\begin{equation}
  \delta p_{l} p_{k}=\left\{
  \begin{array}{@{}ll@{}}
    1, & \text{if}\  cite\left( p_{l} \right)  \geq cite \left( p_{k} \right)  \\
    0, & \text{otherwise}
  \end{array}\right.
\end{equation} 

From Table~\ref{table_example},  $\delta p_{5} p_{4}=1$ and $\delta p_{1} p_{3}=0$ as $cite\left( p_{5}\right) > cite\left( p_{4}\right) $ and $cite\left( p_{3}\right) > cite\left( p_{1}\right) $ correspondingly. 

Given a set $\Theta$ of papers $ \left\lbrace  p_{1}, p_{2}, ... , p_{n} \right\rbrace$ and a community $c_i$, we can now define $w(p_{j})$ as follows. 
\begin{small}
\begin{align} \label{eq_impact}
\begin{split}
w(p_{j}) &= P \left\lbrace X \geq cite \left( p_{j} \right) \right\rbrace \\
& = \frac{\sum_{l=1}^{n} { \Delta c_{i}p_{l} \times \delta p_{l} p_{j}}}{ \sum_{k=1}^{n} \Delta c_{i}p_{k}}
\end{split}
\end{align}
\end{small}

Here  $X= \left\lbrace cite(p_{1}), cite(p_{2}), ... , cite(p_{n}) \right\rbrace$ and  $p_{j} \epsilon \Theta$, where $j = 1, 2, ... , n$. 
  
In Table~\ref{table_example}, community $c_1$ has five papers $(p_1, p_2, p_3, p_4, p_5)$. Among them, three papers $(p_3, p_4, p_5)$ have equal or more citations than paper $p_3$. Therefore, the weight factor of $p_3$ is $w(p_3) = \frac{3}{5}$.
 
Now, the impact factor $imp(p_j)$ of a paper $p_{j}$ can be defined as follows.

\begin{small}
\begin{align} \label{eq_impact}
imp(p_{j}) &= w(p_{j}) \times cite \left( p_{j} \right) 
\end{align}
\end{small}

From Table~\ref{table_example}, the impact of paper $p_3$ can be computed as $imp(p_3) = \frac{3}{5} \times 3 = 1.8$


Based on the above formulation, the influence $I_{c_{i}}$ of community $c_{i}$ is expressed in terms of the \textit{impact} of the papers written by the members of $c_{i}$. Formally, $I_{c_{i}}$ can be defined as follows.

\begin{equation} \label{equ1}
I_{c_{i}} = \frac{\sum_{j=1}^{n} \Delta c_{i} p_{j} \times imp \left( p_{j} \right) \text{,}\ p_{j} \epsilon \Theta}{\sum_{j=1}^{n} \Delta c_{i} p_{j}}  
\end{equation}

We compute the influence of  $c_1$ by using  Equation~\ref{equ1}: $I_{c_{1}} = \frac{0+0+1.8+1.8+10}{5} =  2.72$. Similarly, we can compute influence of  $c_2$ as $I_{c_{2}}= 6.45 $

\subsection{Algorithm}
In this section, we summarize our technique for finding the influential communities. Our proposed algorithm to find the top $K$ influential communities consists of three phases:

In the first phase, we identify the communities $\left\lbrace c_{1}, c_{2}..., c_{n}\right\rbrace $ using \textit{Cluster-Overlap Newman-Girvan Algorithm} (\textit{CONGA}) \cite{conga}.

In the next phase, we compute the influence  $\left\lbrace I_{c_{1}}, I_{c_{2}}, ... , I_{c_{n}}\right\rbrace $ of all identified communities $\left\lbrace c_{1}, c_{2}..., c_{n}\right\rbrace $ by using Equation~\ref{equ1}.

In the final phase, we identify the top $K$ influential communities. We sort $\left\lbrace c_{1}, c_{2}..., c_{n}\right\rbrace $ in a descending order of $\left\lbrace I_{c_{1}}, I_{c_{2}}, ... , I_{c_{n}}\right\rbrace $. Finally, we take the top $K$ communities from the ordered list.

\section{Discussion}
We ran an experiment with a subset of data from DBLP for the duration of the year 2008 to 2016. As a proof of concept, we present a case where h-index fails to measure influence properly. We consider two communities $c_1$ (Aditya G Parameswan, Hector Gracia Molina, Goergia Koutrika, Benjamin Berukovir, Paul Heymann) and $c_2$ (Chih-Jen Lin, Chih-Chung Chang, Chih-Wei Hsu) from database and datamining area. The number of papers written by $c_1$ is $14$ and the citations for those papers are $\lbrace0, 0, 0, 0, 0, 0, 0, 1, 2, 2, 3, 4, 8, 99\rbrace$. However, the number of papers written by $c_2$ is $4$ and the citations for those papers are $\lbrace2, 25, 94, 813\rbrace$.

If h-index is considered as the metric of influence then both $c_1$ and $c_2$ have the same influence, that is 3. However, the papers written by $c_1$ have significantly less number of citations than that of $c_2$. In this case, h-index fails to quantify the influence of these two communities. On the other hand, according to our definition, the influence of $c_1$ and $c_2$ are 11.28 and 67.75, respectively. Therefore, this case proves our stated reasoning behind the definition.

In future, we will extensively analyze the performance of our technique with other competitive influence measures.

\section{Demonstration of VizCom}

We have developed a software, \textbf{VizCom}, an interactive visualization system that enables a user to find the top $K$ \textit{influential} communities from a co-author network. This system also facilitates a user with the overview of these communities along with the details of the authors. The dataset and different visualization modes of \textbf{VizCom} are described below.

\subsection{Dataset}
We collect data set of authors and papers from the Digital Bibliographic Library Browser (DBLP) \cite{dblp}. We further collect the citation data from Google Scholar \cite{gsoc}. After collecting these data, we label all the papers with topics. 
\par 

\subsection{VizCom Modes}
VizCom has five different modes of operation. These modes are \textit{Overview Mode, Community Mode, Focused Community Mode, Author Mode and Focused Author Mode}. A brief description of these modes is given below.
\begin{figure}[!h]
\centering
\frame{\includegraphics[width=3.5in]{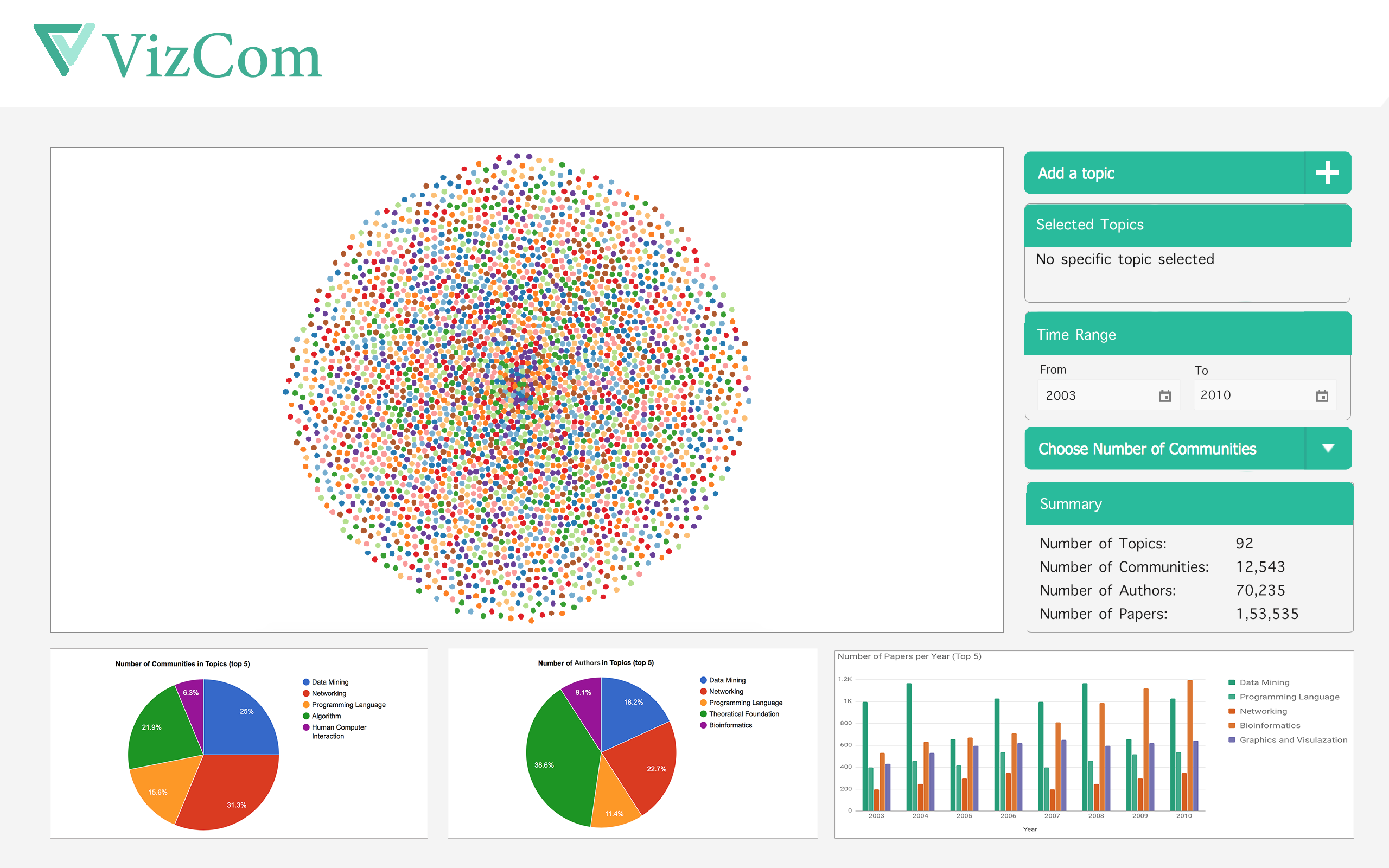}}
\caption{VizCom: \textit{Overview Mode}}
\label{fig_overview}
\vspace{-8mm}
\end{figure}
\subsection{Overview Mode}\label{overview}
In this mode, a user gets an overview of communities of all available topics. In Figure~\ref{fig_overview}, the graphical representation depicts the communities of the topics for a predefined time span. The bottom panel shows the following statistical measures: two pie charts comparing the top five topics with the most number of authors and communities, and a column chart representing the number of papers per year for top five topics with the most number of papers. 

The right panel consists of three input fields and a summarized overview. The three input fields are - \textit{selecting topics, choosing a time span} and \textit{choosing number of communities}. A user can choose variable numbers of topic of interests. This topic selection is a keyword based search. The user can also select a time span, according to his/her requirement. Our system only considers the dataset of this given time span. He/she can also select the number of top $K$ influential communities that he/she wants to focus on; otherwise, all communities are shown. For example, if a user wants to see the top ten influential communities of data mining between 2010 to 2015, he/she has to insert \emph{data mining} as keyword along with the \textit{time span} (2010-2015) and \textit{the number of communities}  as 10. After submitting this information the user will proceed to the \textit{Community Mode}.

\begin{figure}[!h]
\centering
\frame{\includegraphics[width=3.5in]{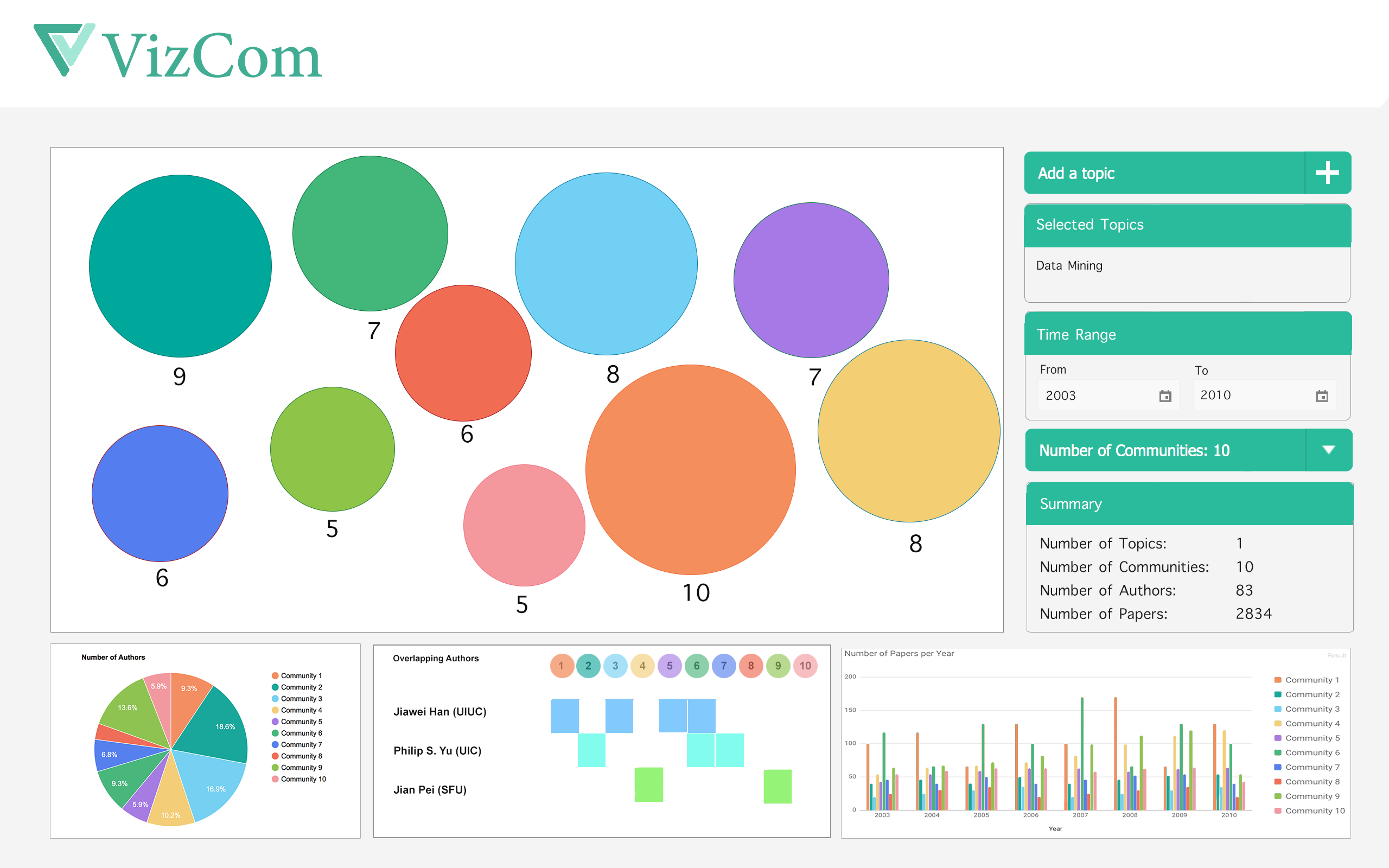}}
\caption{VizCom: \textit{Community Mode}}
\label{fig_community}
\vspace{-8mm}
\end{figure}

\subsection{Community Mode}
After receiving necessary inputs, \textbf{VizCom} presents a user with the top $K$ influential communities along with their normalized influence score. We  normalize the influence score of all communities to get them on a scale of zero to ten. For this, we divide the influence scores of a community with the maximum influence score and then multiply it with 10. Figure~\ref{fig_community} demonstrates \textit{Community Mode} where distinct communities are represented by nodes of different colors. The area of each node characterizes the influence of that community. To illustrate the influence more clearly, the larger nodes denote communities with higher influence. The bottom panel of this mode provides a user with the following details: a pie chart comparing the number of authors per community, a column chart of the number of papers per year for $K$ communities, and a tabular overview of the authors belonging to multiple communities.

A user can select any particular community in order to know more details about that community and and then he/she moves to the \textit{Focused Community Mode}.
\begin{figure}[!h]
\centering
\frame{\includegraphics[width=3.5in]{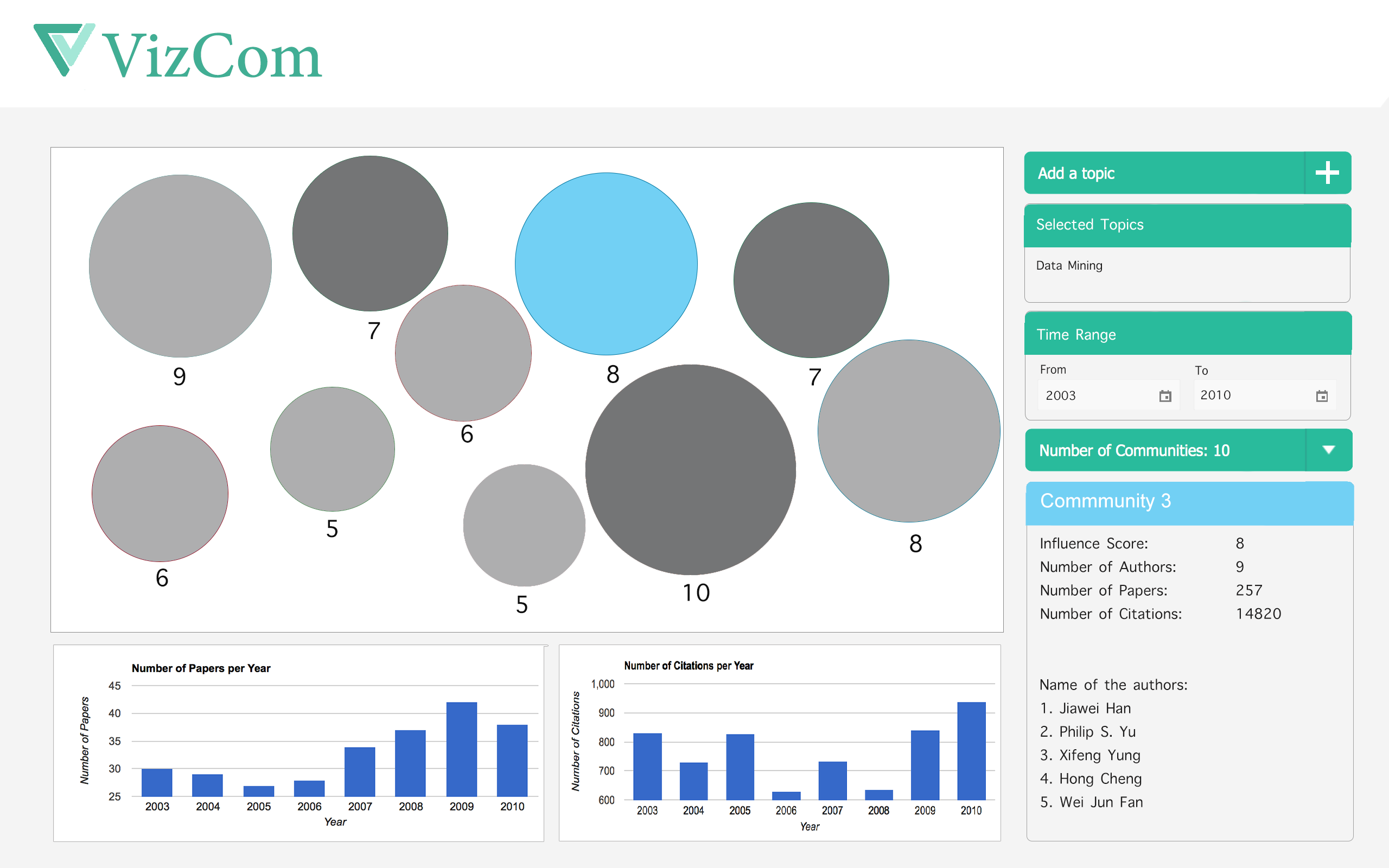}}
\caption{VizCom: \textit{Focused Community Mode}}
\label{fig_focuscommunity}
\vspace{-8mm}
\end{figure}

\subsection{Focused Community Mode}
\textit{Focused Community Mode} of \textbf{VizCom} provides the user extensive information about the selected community. As shown in Figure~\ref{fig_focuscommunity} the right panel shows information about this community along with the input fields mentioned in overview mode. This information includes \textit{community influence, number of authors, number of papers, number of citations, most influential author, overlapping communities, and name of authors.}  

Only the chosen community node preserves its color and the other nodes are transformed into grayscale. Among the other communities, those with overlapping authors with the selected one get darker shades. The bottom panel delivers two charts of the number of papers and citations per year.

\begin{figure}[!h]
\centering
\frame{\includegraphics[width=3.5in]{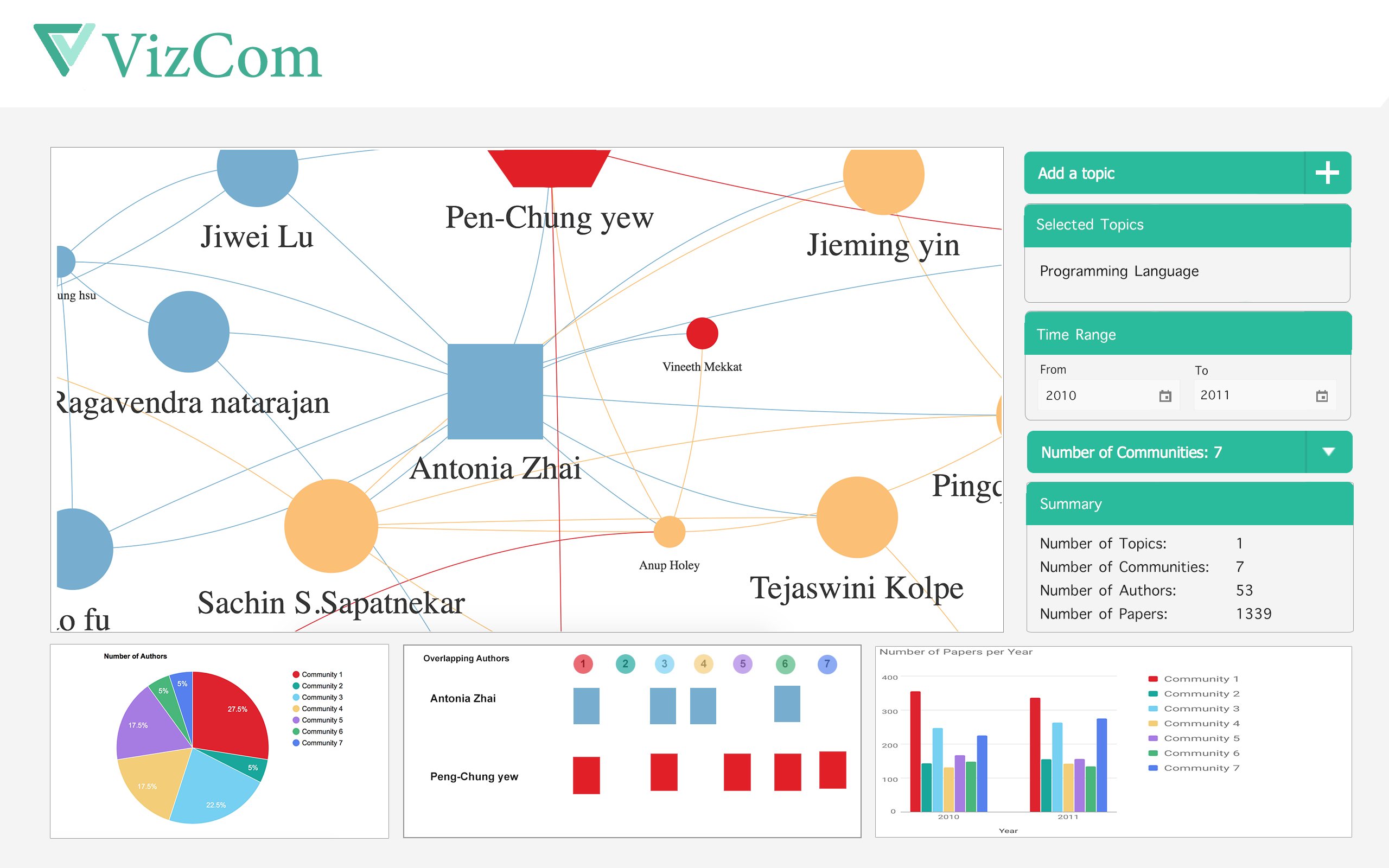}}
\caption{VizCom: \textit{Author Mode}}
\label{fig_author}
\vspace{-3mm}
\end{figure}

\begin{figure}[!h]
\centering
\frame{\includegraphics[width=3.5in]{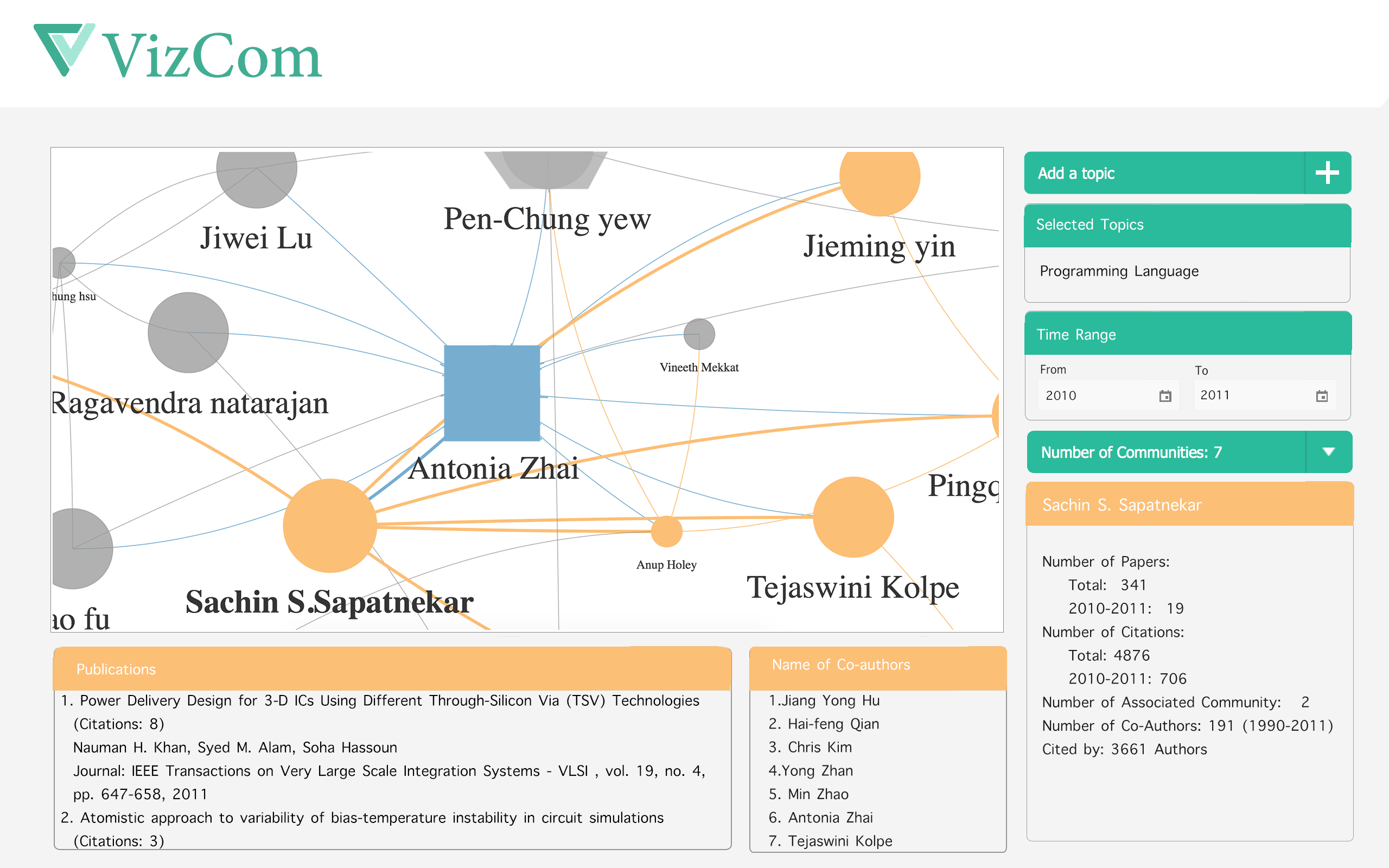}}
\caption{VizCom: \textit{Focused Author Mode}}
\label{fig_authorfocused}
\vspace{-8mm}
\end{figure}

\subsection{Author Mode}
A user can go to \textit{Author Mode} by zooming from \textit{Community Mode} or \textit{Focused Community Mode}. In this mode, the community nodes are expanded. The authors in those communities are shown as nodes and interconnections between them are represented by edges. Moreover, the authors, who belong to multiple communities, are represented as polygons. Here, number of sides of a polygon symbolizes the number of communities. A square representing an author means that this author belongs to four communities. However, we use hexagons to represent the authors belonging to six or more communities. The bottom and the right panels of this mode are similar to the ones of \textit{Community Mode}. If the user selects an author then \textit{Focused Author Mode} is initiated.

\subsection{Focused Author Mode}
This mode focuses on a single author. All the nodes and edges connected with this author are highlighted and relevant information are shown. The information includes \textit{number of papers, number of citations, number of associated communities, number of co-authors, number of authors who cited papers of this author, list of co-authors, and list of publications.}

\section{Conclusion}
In this paper, we define the influence of a community and propose an algorithm to identify the top $K$ influential communities from a network. We also demonstrate a web application to visualize the top $K$ influential communities for a co-author network. Our application enables a user to find the most influential research communities and get detailed information about them. Our work opens a new avenue for a number of potential future works that include incorporating time factor in detecting influence and predicting the future influence of a community.

\subsubsection*{Acknowledgements}
This research is supported by the ICT Division - Government of the People's Republic of Bangladesh.

\bibliographystyle{abbrv}
\bibliography{xbib}  

\begin{thebibliography}{10}

\bibitem{abello}
J.~Abello, M.~G.~C. Resende, and S.~Sudarsky.
\newblock Massive quasi-clique detection.
\newblock In {\em {LATIN}}, pages 598--612, 2002.

\bibitem{dblp}
D.~C. author Network.
\newblock http://dblp.uni-trier.de/xml/.

\bibitem{clauset}
A.~Clauset, M.~E.~J. Newman, and C.~Moore.
\newblock Finding community structure in very large networks.
\newblock {\em Phys. Rev. E}, 70, 2004.

\bibitem{himel}
H.~Dev, M.~E. Ali, and T.~Hashem.
\newblock User interaction based community detection in online social networks.
\newblock In {\em DASFAA}, pages 296--310, 2014.

\bibitem{conga}
S.~Gregory.
\newblock An algorithm to find overlapping community structure in networks.
\newblock In {\em Knowledge Discovery in Databases: {PKDD}}, pages 91--102,
  2007.

\bibitem{hindex}
h~index.
\newblock https://en.wikipedia.org/wiki/H-index.

\bibitem{rong}
R.-H. Li, L.~Qin, J.~X. Yu, and R.~Mao.
\newblock Influential community search in large networks.
\newblock {\em Proc. VLDB Endow.}, 8(5):509--520, 2015.

\bibitem{meo}
P.~D. Meo, E.~Ferrara, G.~Fiumara, and A.~Provetti.
\newblock Generalized louvain method for community detection in large networks.
\newblock {\em CoRR}, abs/1108.1502, 2011.

\bibitem{newman}
M.~Newman.
\newblock Detecting community structure in networks.
\newblock {\em Eur. Phys. J.}, 38(321), 2004.

\bibitem{new}
M.~E.~J. Newman and M.~Girvan.
\newblock Finding and evaluating community structure in networks.
\newblock {\em Phys. Rev. E}, (2):026113.

\bibitem{rag}
U.~N. Raghavan, R.~Albert, and S.~Kumara.
\newblock {Near linear time algorithm to detect community structures in
  large-scale networks}.
\newblock {\em Physical Review E}, 76(3), 2007.

\bibitem{gsoc}
G.~Scholar.
\newblock https://scholar.google.com.

\bibitem{zeng}
Z.~Zeng, J.~Wang, L.~Zhou, and G.~Karypis.
\newblock Out-of-core coherent closed quasi-clique mining from large dense
  graph databases.
\newblock {\em {ACM} Trans. Database Syst.}, 32(2):13, 2007.

\end{thebibliography}

\end{document}